\documentclass[aps,prl,reprint,groupedaddress]{revtex4-1}

\usepackage{graphicx}
\usepackage{amsmath} 
\usepackage{epstopdf}
\usepackage{amsmath} 
\usepackage{color}

\newcommand{\ndg}{{\phantom{\dagger}}}
\newcommand{\dg}{\dagger}
\newcommand{\ket}[1]{\left|#1\right\rangle}
\newcommand{\bra}[1]{\left\langle #1\right|}

\newcommand{\fss}{\delta_\text{FSS}}

\begin{document}


\title{Entanglement robustness to excitonic spin precession in a quantum dot}


\author{Samir Bounouar*$^1$, Gabriel Rein$^1$, Kisa Barkemeyer$^2$, Julian Schleibner$^2$, Peter Schnauber$^1$, Manuel Gschrey$^1$, Jan-Hindrik Schulze$^1$, Andr\'{e} Strittmatter$^1$, Sven Rodt$^1$, Andreas Knorr$^2$, Alexander Carmele$^2$, and Stephan Reitzenstein$^1$}
\affiliation{ \em $^1$Institut f\"ur Festk\"orperphysik, Technische Universit\"at Berlin, 10623 Berlin, Germany\\ $^2$Institut f\"ur Theoretische Physik, Technische Universit\"at Berlin, 10623 Berlin, Germany }


\date{\today}

\begin{abstract}
A semiconductor quantum dot (QD) is an attractive resource to generate polarization-entangled photon pairs. We study the excitonic spin precession (flip-flop) in a family of QDs with different excitonic fine-structure splitting (FSS) and its impact on the entanglement of photons generated from the excitonic-biexcitonic radiative cascade. Our results reveal that coherent processes leave the time post-selected entanglement of QDs with finite FSS unaffected while changing the eigenstates of the system. The flip-flop's precession is observed via quantum tomography through anomalous oscillations of the coincidences in the rectilinear basis. A theoretical model is constructed with the inclusion of an excitonic flip-flop rate and is compared with a two-photon quantum tomography measurement on a QD exhibiting the spin flip-flop mechanism. A generalization of the theoretical model allows estimating the degree of entanglement as a function of the FSS and the spin-flip rate. For a finite temporal resolution, the negativity is found to be oscillating with respect to both the FSS and the spin-flip rate. This oscillatory behavior disappears for perfect temporal resolution and maximal entanglement is retrieved despite the flip-flop process.

\end{abstract}


\maketitle

\section{Introduction}

Quantum mechanical entanglement has proven to be a crucial prerequisite for experimental realizations of quantum communication protocols~\cite{knill,kimble}, quantum computing~\cite{pan} or fundamental tests of quantum mechanics~\cite{wheis}. In this context, self-assembled semiconductor quantum dots (QDs) are excellent candidates for the on-demand generation of polarization-entangled photon pairs via the radiative exciton-biexciton cascade \cite{benson}. An excitonic fine-structure splitting (FSS) below the radiative linewidth has been considered as a prerequisite for the realization of such sources of entangled photon pairs~\cite{akopian, steven,ondem}. More recently, time post-selection of the excitonic wave packet has been applied as a which-path erasure for the generation of polarization-entangled photon pairs from QDs with significant FSS~\cite{winik, bounouarapl}. Since it requires no technologically-demanding suppression of the FSS, the latter technique has proven to be very practical  for the realization of complex quantum communication schemes~\cite{Zopf} and is still unavoidable in the telecom wavelength range~\cite{huwer}. 
Polarization entanglement is robust to pure dephasing when decoherence affects the excitonic (X) and the biexcitonic (XX) phase in a same manner~\cite{cohsteven,carmele2011analytical,reiter2019distinctive,seidelmann2019strong,hohenester2007phonon}. This is the case for standard electron-phonon interaction in type III-V heterostructures and can be furthermore distilled via spectral filtering \cite{del2013distilling}. However, spin-flip induced cross-dephasing processes still threaten the symmetry-imposed erasure of the which-path information and are believed to be the main obstacle to the generation of entanglement with unity fidelity~\cite{trotta}. 
This is in particular true if the whole excitonic wave packet is collected in a time-integrated quantum state tomography setup~\cite{cohsteven,akopian,steven,ondem,carmele2010formation}. If in contrast post-selection is applied, the which-path information is again lost and maximum entanglement is established \cite{bounouarapl,versteegh,winik}. This conservation of entanglement is usually overlooked by experimentalists since a pre-selection of QDs showing no spin flip is applied before the tomography and candidates which have otherwise advantageous properties for the generation of polarization-entangled photons may be omitted. For example, a QD exhibiting a small spin-flip rate is to be preferred against a QD without spin flips but large FSS.
In this work, we present a method for predicting the expected entanglement quality with the negativity as a measure for the degree of entanglement \cite{verstraete2001comparison,lee2003convex,PhysRevLett.77.1413} as a function of the measured spin-flip rates for a set of different QDs, considering a time post-selection of the excitonic wave packet.
In the derivation we make use of the Wigner-Weisskopf approximation assuming the dynamics to be Markovian. Beyond this limit, non-Markovianity could yield interesting effects such as the possibility to control the degree of entanglement via coherent time-delayed feedback \cite{Barkemeyer2019, Carmele2019, Strauss2019, Hein2014}.
Our method is tested on an InGaAs QD deterministically integrated into a microlens showing a non-null spin-flip rate. It is then generalized by deriving the negativity as a function of the FSS and the spin-flip rate analytically which is found to be oscillating for both parameters with a frequency defined by the actual experimental time resolution.

Interestingly, the precession of the excitonic spin configuration can be observed in the tomography measurements in the linear basis. In the presence of such spin precession, we find that the post-selected entanglement quality is solely limited by the experimental resolution. In our case, the interaction with nuclear spins results in a unitary transformation, which is non-dissipative and thus no obstacle to the time post-selected generation of unity-entanglement.

\section{Sample and setup}
Our experiments are carried out on self-assembled InGaAs/GaAs QDs grown by metal-organic chemical vapor deposition. The QDs are integrated deterministically into microlenses with a backside AlGaAs/GaAs distributed Bragg reflector by 3D in-situ electron-beam lithography~\cite{gschrey}. The microlens sample is placed in a Helium flow cryostat and operated at a temperature of $5$~K. Micro-photoluminescence ($\mu$PL) measurements are performed with a standard cross-correlation spectroscopy setup, where detectors are silicon avalanche photodiode based single-photon counting modules (SPCMs) with a temporal resolution of $550$~ps (FWHM). A quantum tomography is realized via superconducting nanowire single photon detectors (SNSPDs) with a temporal resolution of $100$~ps (FWHM). The studied QDs are excited with a continuous-wave 780 nm laser diode.

\begin{figure}[htbp]
\centerline{\includegraphics[width=\linewidth]{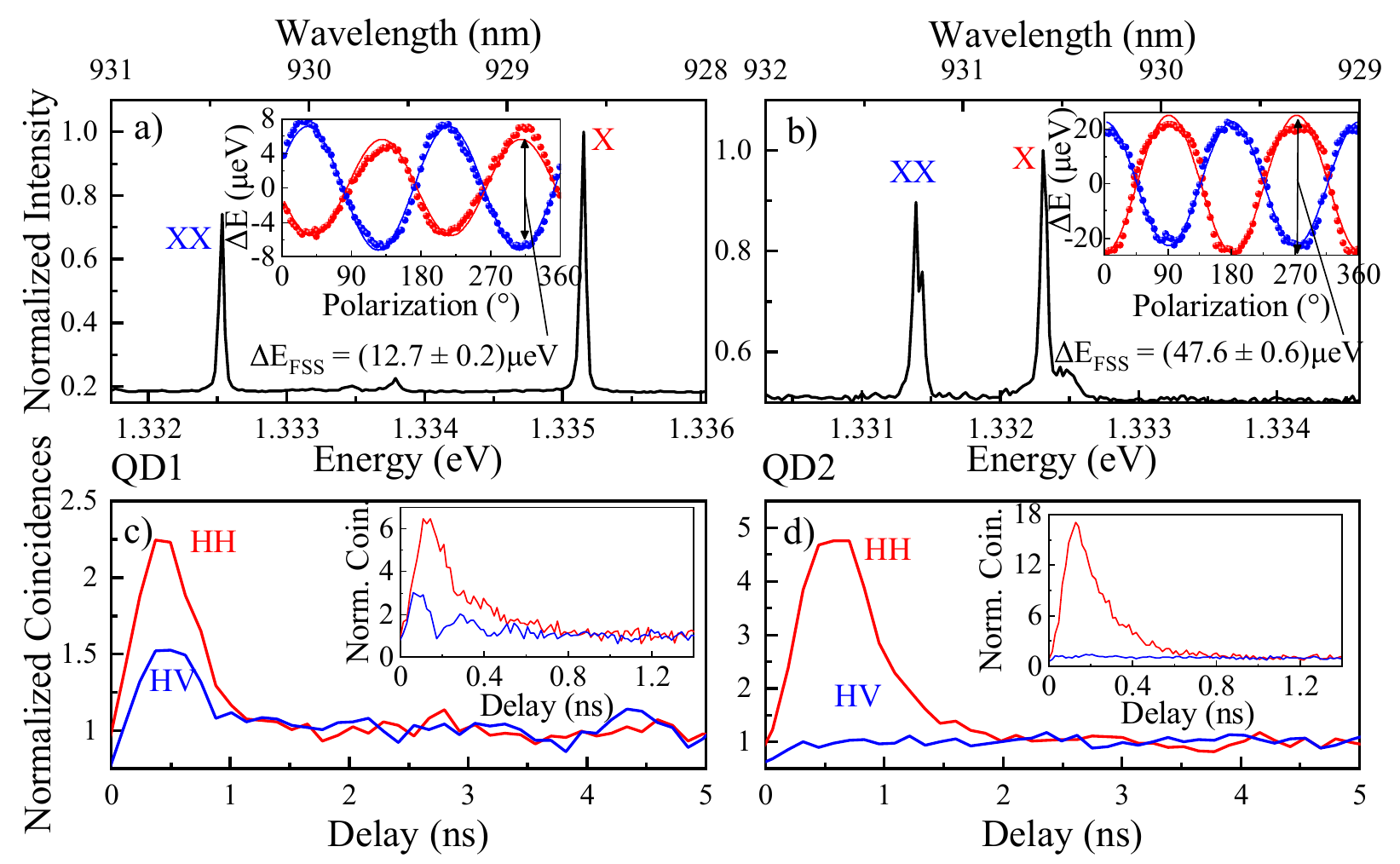}}
\caption{\footnotesize Typical $\mu$PL emission spectra of the studied quantum dots QD1 with a FSS of $\fss = 12.7 \pm 0.2~\mu$eV  (a) and QD2 with a FSS of $\fss = 47.6 \pm 0.6~\mu$eV (b). Polarization-resolved measurements of the FSS are displayed in the inset of each graph. Polarization-resolved cross-correlation measurement of QD1 (c) and QD2 (d) in the HH basis (red) and in the HV basis (blue) obtained with SPCMs with 550 ps temporal resolution. The inset shows the same measurement using SNSPDs with a temporal resolution of 100 ps for comparison.}
\label{fig:fig1}
\end{figure}

\section{Spin-flip rate estimation}

\begin{figure}[htbp]
\flushleft
\includegraphics[width=0.9\linewidth]{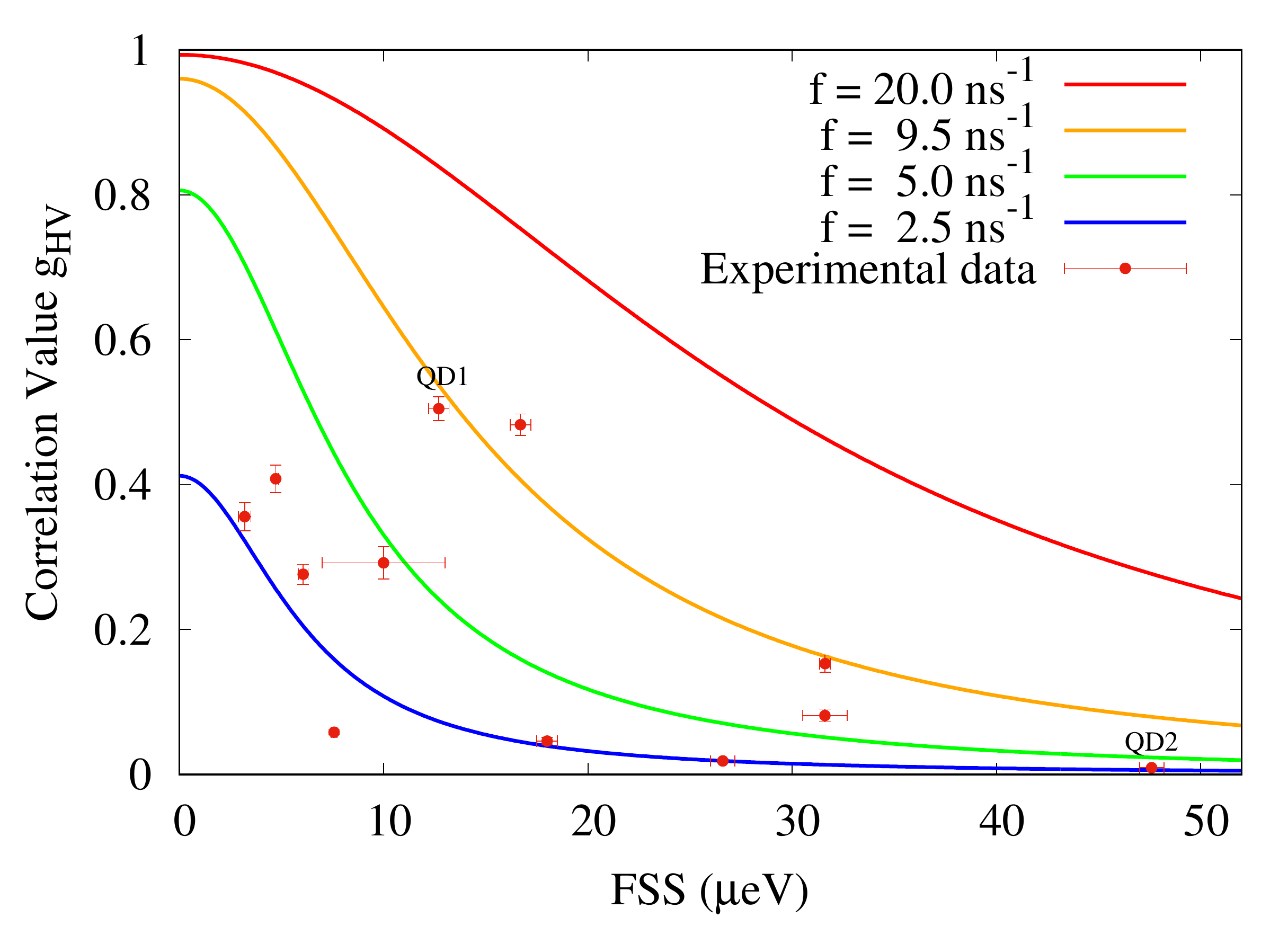}
\caption{\footnotesize Measured correlation values $g_{HV}$ for several QDs as a function of their FSS. The two QDs discussed in the text are indicated as QD1 ($\fss = 12.7 \pm 0.2~\mu$eV) and QD2 ($\fss=47.6 \pm 0.6~\mu$eV). The colored curves are extracted from the theoretical model for different spin-flip rates. }
\label{fig:fig2}
\end{figure}

Fig.~\ref{fig:fig1} shows $\mu$PL spectra of two different QDs, labeled QD1 (Fig.~\ref{fig:fig1}(a)) and QD2 (Fig.~\ref{fig:fig1}(b)), with a FSS of $\fss=12.7 \pm 0.2~\mu$eV and $\fss=47.6 \pm 0.6~\mu$eV, respectively, as displayed in the insets. The spectra were obtained under non-resonant (780 nm) continuous-wave excitation. For both QDs, the two characteristic emission lines result from the radiative recombination of the exciton (X) and the biexciton (XX) states. 
Excitonic spin-flip rates can be estimated by measuring the XX and X photon correlations in the horizontal and the vertical basis where the basis states are set as aligned along the QD main axes. Fig.~\ref{fig:fig1}(c) and (d) show the normalized coincidences measured between the XX and X photons with the XX photon triggering the correlation measurement. 
Photon bunching is not only found as expected in the horizontal-horizontal ($HH$) basis (in red) but also in the horizontal-vertical ($HV$) basis (in blue). This observation reveals the possibility for the excitonic population to transit from one excitonic component to the other, preventing the QD from emitting two co-polarized photons which would be impossible without spin flips. The insets of Fig.~\ref{fig:fig1}~(c) and (d) display the same polarization-resolved correlation measurement done with the superior resolution of the SNSPDs ($100$~ps) for QD1 and QD2, respectively. In the case of QD1, the spin-flip related oscillations become visible for the $HV$ correlations and a shoulder adds to the bunching observed for the $HH$ correlations. These features are caused by a precession of the excitonic spins about the horizontal axis. This is very similar to a Larmor precession about an effective magnetic field created by the surrounding nuclear spins~\cite{overhauser}. The latter can be provided by the anisotropic nuclear spin distribution surrounding the QD~\cite{hinz}. It has been shown that the strength of the resulting effective magnetic field necessary for such a precession is reachable in QDs~\cite{maletinsky}. The exchange interaction between the electron and hole excitonic spins can also lead to their relaxation through a mixing of the two excitonic components and could lead to the observed spin-flip effect~\cite{maialle,Nick}.
By comparing both bunchings, one can evaluate the excitonic spin-flip rates. In the following, the spin-flip rate will be studied by considering the ``correlation value" $g_{HV}$ defined as the ratio between the two bunchings observed for the cross-correlations in the $HH$ basis and in the $HV$ basis after subtraction of the Poisson level: $g_{HV}=A_{HV}/A_{HH}$ with $A_{HV}$ ($A_{HH}$) the time-integrated $HV$ ($HH$) coincidences for positive delays (see  Fig.~\ref{fig:fig1}(c) and (d). The correlation value $g_{HV}$ is zero when no spin flips occur and reaches 1 when the spin-flip rate is much larger than the excitonic decay rate. Fig.~\ref{fig:fig2} shows the different $g_{HV}$ values evaluated for several QDs with different FSS. The two extreme situations in our family of QDs are represented by the two QDs marked as QD1 (highest correlation value) and QD2 (lowest correlation value). The insets of  Fig.~\ref{fig:fig1} (c) and (d) show the corresponding correlation functions for these two QDs. The curves result from the theoretical model (described below) for different spin-flip rates. They show that for a given spin-flip rate the correlation value is decreasing with the FSS and allow an estimation of the occurring spin-flip rate. The closest curve(represented in orange) to QD1 corresponds to a spin-flip rate of $f=9.5$~ns$^{-1}$, which is consistent with the fit parameter used for the fitting of the quantum tomography. 

\section{Quantum tomography and negativity}

To be able to understand the impact of the spin-flip rate on the post-selected entanglement, we develop a theoretical approach:
We model the system in the Schr\"odinger picture and take into account a spin-flip rate $f$ and a FSS $\fss$ within a biexciton-cascade process.
We show that the case with spin flips and FSS is isomorphic to the case without spin flips but with FSS.
Therefore, in the case of a time-resolved quantum state tomography setup, the spin-flip rate has no detrimental effect if the temporal resolution is high in relation to spin-flip rate and FSS.
The loss of which-path information is preserved under this particular unitary evolution and we can still find a high degree of entanglement.
In order to show this, we map the biexciton-cascade process with finite spin-flip rate and FSS to the isomorphic case without spin flips.
This is done via the diagonalization of the Hamiltonian.
The Hamiltonian in the interaction picture reads
\begin{align}
H/\hbar &= 
\frac{\fss}{2} ( \sigma_{VV} - \sigma_{HH} )
+ f  ( \sigma_{VH} + \sigma_{HV} )  \notag\\ \notag
&+ g_0
\int \text{d}\omega 
\left( 
a^\dg_{\omega,BH}
\sigma_{HB}
e^{i(\omega-\omega_{BX})t}
 \right.\\& \qquad \quad \quad \left.
 +
a^\dg_{\omega,BV}
\sigma_{VB}
e^{i(\omega-\omega_{BX})t}
+ \text{h.c.}
\right) \notag\\ \notag
&+ g_0
\int \text{d}\omega 
\left( 
a^\dg_{\omega,XH}
\sigma_{GH}
e^{i(\omega-\omega_{XG})t}
\right. \\ &\qquad \quad \quad \left. +
a^\dg_{\omega,XV}
\sigma_{GV}
e^{i(\omega-\omega_{XG})t}
+ \text{h.c.}
\right) 
\end{align}
where we spectrally distinguish the biexciton ($B$) from the exciton photons ($X$) with horizontal ($H$) or vertical ($V$) polarization in frequency mode $\omega$ which are annihilated (created) via $a^{(\dg)}_{\omega, B/X H/V}$.
The photon operators satisfy the bosonic commutation relation
$ [a^\ndg_{\omega,im},a^\dg_{\omega',jn}]=\delta(\omega-\omega')
\delta_{ij}\delta_{mn}$, $i,j \in \{B,X \}$, $m,n \in \{H,V\}$.
Transitions between the electronic states are described by the flip operators $\sigma_{ij}=\ket{i}\hspace{-.1cm}\bra{j}$, $ i,j \in \{B, X_H,X_V, G\}$ with $\ket{B}$ denoting the biexciton, $\ket{X_H}$ the horizontally polarized exciton, $\ket{X_V}$ the vertically polarized exciton, and $\ket{G}$ the ground state. For reasons of readability we replace the label of the excitonic states in the subscript by the respective direction of polarization $X_H \rightarrow H$, $X_V \rightarrow V$.
The exciton energy $\hbar \omega_{XG}$ is centered between the energy of the horizontally polarized exciton state $\hbar \omega^H_{XG}=\hbar \left(\omega_{XG}-\fss/2\right)$ and the one of the vertically polarized exciton state $\hbar \omega^V_{XG}=\hbar \left( \omega_{XG}+\fss/2\right)$. The biexciton-exciton transition energy is $\hbar \omega_{BX}=\hbar \left(\omega_{XG}-\omega_\text{bind}\right)$ where $\hbar\omega_\text{bind}$ describes the binding energy of the biexciton.
The electron-photon interaction strength $ g_0 $ is related to the radiative decay constant $\Gamma$ via $ g_0=\sqrt{\Gamma}/\pi$.
We first diagonalize the spin-flip part of 
the Hamiltonian via introducing new excitonic basis states
\begin{align}
\ket{X_-} =& \alpha \ket{X_H} - \beta \ket{X_V}, \\
\ket{X_+} =& \beta \ket{X_H} + \alpha \ket{X_V} 
\end{align}
with
$ \alpha^2=(1+\fss/(2\Omega))/2$, $ \beta^2=(1-\fss/(2\Omega))/2$
and the renormalized frequency $ \Omega^2=f^2+(\fss/2)^2$.
This transformation leads to a Hamiltonian which is diagonal in the spin-flip dynamics
\begin{align}
H/\hbar &= 
\Omega ( \sigma_{++} - \sigma_{--}) \notag \\ \notag
&+ g_0
\int \text{d}\omega 
\left[ 
\sigma_{-B}
\left(\alpha a^\dg_{\omega,BH}
- \beta a^\dg_{\omega,BV}\right)
e^{i(\omega-\omega_{BX})t}
\right. \\ \notag & \left. \qquad +
\sigma_{+B}
\left(\beta a^\dg_{\omega,BH}
+ \alpha a^\dg_{\omega,BV}
\right)
e^{i(\omega-\omega_{BX})t}
+ \text{h.c.}
\right] \\ 
&+ g_0
\int \text{d}\omega 
\left[
\sigma_{G-}
\left(\alpha a^\dg_{\omega,XH}
- \beta a^\dg_{\omega,XV}\right)
e^{i(\omega-\omega_{XG})t}
\right. \notag  \\ & \left. \qquad +
\sigma_{G+}
\left(\beta a^\dg_{\omega,XH}
+ \alpha a^\dg_{\omega,XV}
\right)
e^{i(\omega-\omega_{XG})t}
+ \text{h.c.}
\right].
\end{align}
We furthermore introduce new photonic operators
\begin{align}
&a^\dg_{\omega,B/X +} = 
\beta a^\dg_{\omega,B/XH}
+ \alpha a^\dg_{\omega,B/XV}, \\
&a^\dg_{\omega,B/X -} = 
\alpha a^\dg_{\omega,B/XH}
- \beta a^\dg_{\omega,B/XV}.
\end{align}
Due to the definitions of $\alpha$ and $\beta$, these
operators are bosonic operators and thus satisfy the commutation relation 
$[a^\ndg_{\omega,im},a^\dg_{\omega',jn}]=\delta(\omega-\omega')
\delta_{ij}\delta_{mn}$, $i,j \in \{B,X\}$, $m,n \in \{+,- \}$.
This can easily be shown by using $\alpha^2+\beta^2=1$ and the commutation relations of the photon operators in the $H/V$ basis.
Plugging in the new photonic operators, the Hamiltonian reads
\begin{align}
H/\hbar &= 
\Omega ( \sigma_{++} - \sigma_{--})  \notag\\ \notag
&+ g_0
\int \text{d}\omega 
\left( 
\sigma_{-B}
a^\dg_{\omega,B-}
e^{i(\omega-\omega_{BX})t}
\right. \\ &\left. \qquad \quad \quad \quad \notag+
\sigma_{+B}
a^\dg_{\omega,B+}
e^{i(\omega-\omega_{BX})t}
+ \text{h.c.}
\right) \\ 
&+ g_0
\int \text{d}\omega 
\left( 
\sigma_{G-}
a^\dg_{\omega,X-}
e^{i(\omega-\omega_{XG})t} \notag
\right. \\ &\left. \qquad \quad \quad \quad +
\sigma_{G+}
a^\dg_{\omega,X+}
e^{i(\omega-\omega_{XG})t}
+ \text{h.c.}
\right)
\end{align}
which is isomorphic to the Hamiltonian without spin flips.
Next, we calculate the state of the system. Assuming the system to initially be in the biexciton state, $\ket{\Psi(0)} = \ket{B, \text{vac}}$,
the normalized wave function reads
\begin{align}
&\ket{\Psi(t)}
= e^{-2\Gamma t} \ket{B,\text{vac}} - i\sqrt{2\Gamma} e^{-\Gamma t}
\int_0^t \text{d}t'
e^{-\Gamma t'} 
 \notag \\ &\times \left[
e^{-i\Omega (t-t')} \ket{X_+,+(t')} \notag
 +
e^{i\Omega(t-t')} \ket{X_-,-(t')}
\right] \notag \\
& - 2\Gamma 
\int_0^t \text{d}t'
\int_0^{t'} \text{d}t''
e^{-\Gamma (t'+t'')}
 \notag \\ &\times \left[
e^{-i\Omega (t'-t'')} \ket{+(t''),+(t')}
+
e^{i\Omega (t'-t'')} \ket{-(t''),-(t')}
\right].
\label{two-ph-wf}
\end{align}
Here $\ket{B, \text{vac}}$ describes the system in the biexciton state and no photons in the reservoir, $\ket{X_{\pm},\pm(t)}$ refers to the system being in the $\pm$ exciton state after the emission of a $\pm$ biexciton photon
where we have defined the biexcitonic photon state as $\ket{\pm(t)}_{B}=(2\pi)^{-1/2}\int d\omega \exp[i(\omega -\omega_{BX}) t] a^\dg_{\omega,\pm}\ket{\text{vac}}$. The state $\ket{\pm(t'), \pm(t)}$ describes a $\pm$ biexciton photon and a $\pm$ exciton photon. The excitonic photon state is defined analogously as $\ket{\pm(t)}_{X}=(2\pi)^{-1/2}\int d\omega \exp[i(\omega -\omega_{XG}) t] a^\dg_{\omega,\pm}\ket{\text{vac}}$.
Looking at the above result we see that spin-flips do not change the physics of the biexciton cascade qualitatively as the solution is isomorphic to the case without spin flips but with finite FSS.
In an integrated quantum state tomography, spin flips are detrimental to maximal entanglement just as an FSS is.
For a time-resolved quantum state tomography, however, with a resolution high enough in relation to spin-flip rate and FSS, entanglement is preserved since which-path information is erased. 
Using the state derived above, the time-resolved quantum state tomography can be calculated.
In the following, we assume that the cascade process is finished.
In this limit, the two-photon wave function reads in the basis of horizontally and vertically polarized photons
\begin{align}
&\ket{\Psi(\infty)}
= - 2\Gamma \notag
\int_0^\infty \text{d}t'
\int_0^{t'} \text{d}t''
e^{-\Gamma (t'+t'')} \\ & \qquad \times
\left(
\alpha^2 e^{i\Omega (t'-t'')} 
+
\beta^2 e^{-i\Omega (t'-t'')}
\right)\ket{H(t''),H(t')}\notag \\
&- 2\Gamma\alpha\beta 
\int_0^\infty \text{d}t'
\int_0^{t'} \text{d}t''
e^{-\Gamma (t'+t'')} \notag \\
& \qquad  \times \left(
e^{-i\Omega (t'-t'')}
-e^{i\Omega (t'-t'')} 
\right)\ket{H(t''),V(t')} \notag \\
&- 2\Gamma\alpha\beta 
\int_0^\infty \text{d}t'
\int_0^{t'} \text{d}t''
e^{-\Gamma (t'+t'')} \notag \\
& \qquad \times 
\left( 
e^{-i\Omega (t'-t'')} 
-
e^{i\Omega (t'-t'')}
\right)\ket{V(t''),H(t')} \notag\\
&- 2\Gamma 
\int_0^\infty \text{d}t'
\int_0^{t'} \text{d}t''
e^{-\Gamma (t'+t'')} \notag \\
& \qquad  \times \left(
\alpha^2 e^{-i\Omega (t'-t'')} 
+
\beta^2 e^{i\Omega (t'-t'')}
\right)\ket{V(t''),V(t')}.
\end{align}
We can rewrite the wave function as
\begin{align}
\ket{\Psi(\infty)}
= &
\int_0^\infty \text{d}t'
\int_0^{t'} \text{d}t''
\left[
P_{D}(t'',t') \ket{H(t''),H(t')}  \notag
 \right. \\ & \qquad \qquad \qquad \notag \left. +
P^*_{D}(t'',t') \ket{V(t''),V(t')} \right] \notag\\
+ & \int_0^\infty \text{d}t'
\int_0^{t'} \text{d}t''
\left[
P_{ND}(t'',t') \ket{H(t''),V(t')} \notag
 \right. \\ & \qquad \qquad \qquad \left. -
P^*_{ND}(t'',t')
\ket{V(t''),H(t')}\right]
\end{align}
with
\begin{align}
&P_D(t'',t') =  
- 2\Gamma 
e^{-\Gamma (t'+t'')} \notag \\
&\qquad \times \Big\{
\cos[\Omega (t'-t'')] 
+i\frac{\fss}{2\Omega}
\sin[\Omega (t'-t'')] \Big\}, \\
& P_{ND}(t'',t') =  
i 2\Gamma\frac{f}{\Omega} 
e^{-\Gamma (t'+t'')}
\sin[\Omega (t'-t'')].
\end{align}
The detection of a biexciton (B) or exciton (X) photon with polarization $H/V$ at time $t_D$ can be described via the operators
\begin{align}
D^{(+)}_{B \hspace{.15em}H/V}(t_D)&= 
\frac{1}{\sqrt{2\pi}}
\int \text{d}\omega e^{-i(\omega - \omega_{BX})  t_D} a_{\omega,B \hspace{.15em}H/V}, \\
D^{(+)}_{X \hspace{.15em}H/V}(t_D)&= 
\frac{1}{\sqrt{2\pi}}
\int \text{d}\omega e^{-i(\omega - \omega_{XG})  t_D} a_{\omega,X \hspace{.15em}H/V}.
\end{align}
Using the relation
\begin{multline}
    D^{(+)}_{X H}(t_X) D^{(+)}_{B H}(t_B)   \ket{H(t'') H(t')} \\ =\delta(t_X -t'')\delta(t_B-t')\ket{\text{vac}}
\end{multline}
and analogous relations for the other polarization directions, we can describe the detection of the exciton photon at $t_X$ and the detection of the biexciton photon at $t_B$ in the basis $\lbrace \ket{HH},\ket{HV},\ket{VH},\ket{VV} \rbrace$ as
\begin{multline}
\ket{\Psi(t_X,t_B)}= \\
\left(P_{D}(t_X,t_B), P_{ND}(t_X,t_B),-
P^*_{ND}(t_X,t_B), P^*_{D}(t_X,t_B) \right)^T.
\end{multline}
The corresponding measurement matrix can be constructed via
\begin{align}
\rho_{im,jn}(t_X,t_B)=\langle i \hspace{.15em}m|\Psi(t_X, t_B)
\rangle\langle \Psi(t_X, t_B)| j \hspace{.15em}n \rangle   
\end{align}
with $i,m, j, n \in \{H,V \}$. Note that if we normalize the above state and matrix, the conditional probability to measure the photon pair in certain directions of polarization given that we measure them at time $t_X$ and $t_B$ can be calculated from their elements.
Convoluting the elements of the measurement matrix with the system response function we can fit the experimentally determined tomography.

\begin{figure}[b!]
\centerline{\includegraphics[scale=0.45]{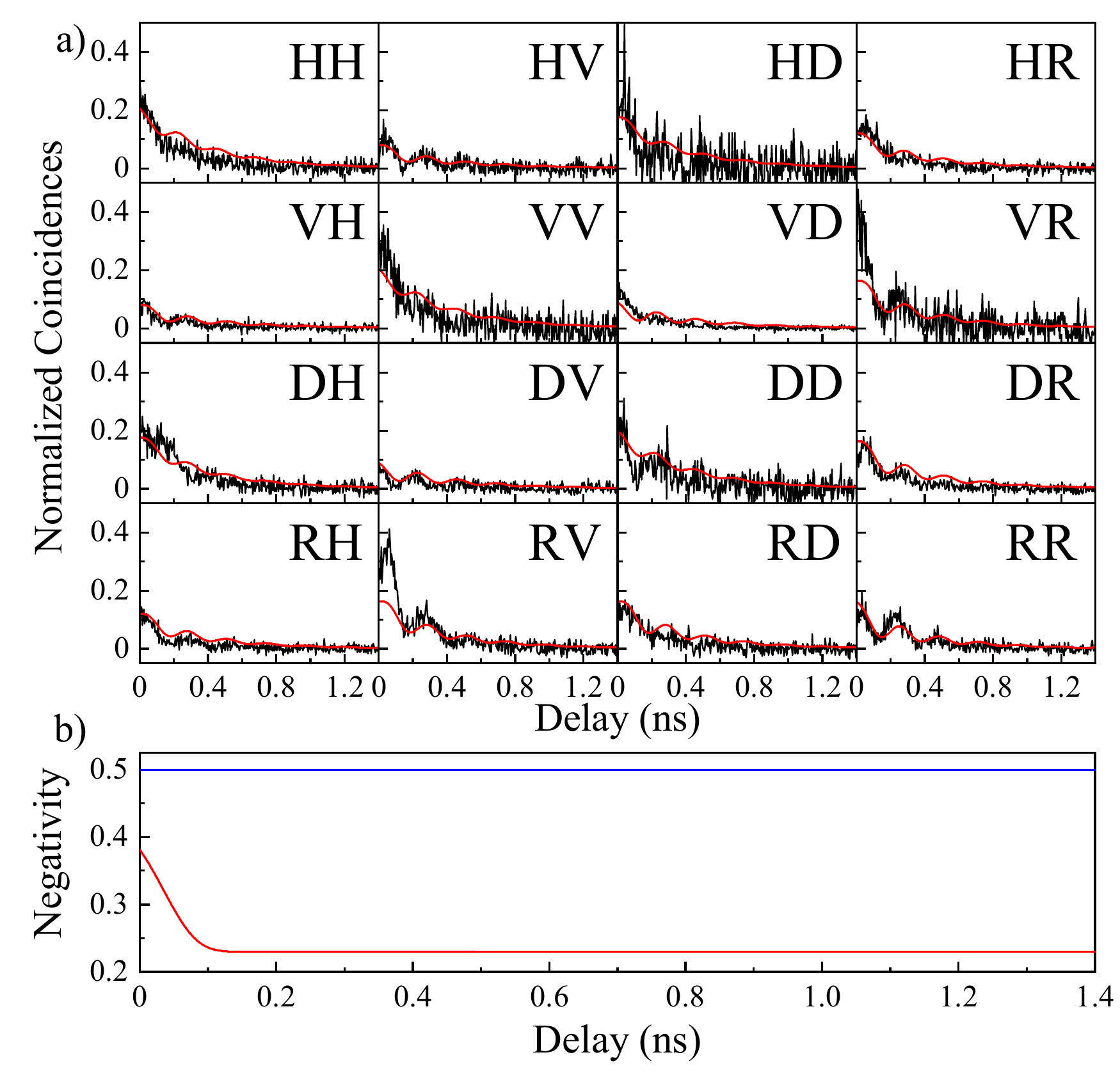}}
\caption{\footnotesize (a) 16 polarization-resolved measurements representing the full tomography of QD1. The red curves result from the fitting model taking the spin-flip processes in the QD into account and convoluted to the setup resolution. This fitting is used to calibrate the model and to extrapolate it for the other QDs studied here. (b) Extrapolated negativity as a function of the delay for a perfect temporal resolution (blue) and a $100$ ps resolution (red). }
\label{fig:fig3}
\end{figure}
Fig.~\ref{fig:fig3}(a) shows the 16 polarization cross-correlation measurements of QD1 necessary for the full tomography \cite{James2001}. The black lines represent the experimental data, the red lines are obtained from the analytical model using the experimental parameters $\Gamma=2.4$~ns$^{-1}$ and $\fss=12.7~\mu$eV with an estimate of $f=9.5$~ns$^{-1}$ for the spin-flip rate. Note that pure dephasing processes are not considered as they do not affect the entanglement\cite{cohsteven,carmele2011analytical,reiter2019distinctive,seidelmann2019strong,hohenester2007phonon}. Overall, we obtain a very good agreement between experiment and theory. However, for some elements of the two-photon density matrix (e.g. $RV$ and $VR$), the bunching values measured in the $HV$ and $VH$ correlations are underestimated by the fits. This discrepancy can be explained by some unintentional fluctuations of the excitation power which is influencing the bunching amplitude of the correlations.
Fig.~\ref{fig:fig3}(b) shows the numerically obtained negativity $\mathcal{N}$ as a function of the delay. The negativity is a measure of the degree of entanglement of the two photons. It is defined as the absolute value of the negative eigenvalue of the partially transposed two-photon density matrix \cite{PhysRevLett.77.1413}. A value of $\mathcal{N}>0$ signifies an entangled two-photon state with maximal entanglement at $\mathcal{N}=1/2$.
If perfect time resolution is assumed, the negativity is maximal (i.e. $\mathcal{N} = 1/2$) independent of the delay since the two-photon wave function remains maximally entangled, cf. Eq.~\eqref{two-ph-wf}. Spin-flip processes and FSS do not change the symmetry of the problem if no time-average is applied. As a result, the negativity, as a basis-independent measure for 
the degree of polarization entanglement, is not changed whether the system is expressed in the  $\{H,V\}$ or in the $\{+,-\}$ basis. However, for finite time-resolution the spin-flip and the FSS dynamics result in a reduced negativity. If the photon-detection events are not post-selected but integrated and therefore averaged, spin-flip dynamics have a detrimental influence. In the case of temporal post-selection for non-zero FSS oscillations corresponding to the precession of the excitonic phase are observed in the circular $(RR)$ and in the diagonal basis $(DD)$ as is also the case in the absence of spin flips. Moreover, due to the spin-flip process oscillations can also be observed in the rectilinear basis $(VH$, $HV$, $VV$ and $HH)$. Such oscillations are a direct consequence of the excitonic spin flips and are not observed in the case of spin-flip rate zero. From these data and the theory, one can at the same time test and calibrate the model and evaluate the quality of the emitted entanglement by this QD (QD1) in terms of the negativity \cite{verstraete2001comparison,lee2003convex,PhysRevLett.77.1413}. In the following, we extrapolate the model to the general evaluation of the entanglement as a function of the FSS and the spin-flip rate for any QD.

To evaluate the degree of entanglement of the two photons theoretically, we look at the normalized density matrix $\rho_N = \rho/\text{tr}(\rho)$ and calculate its negativity. If we assume a perfect resolution of the detection process, we obtain a negativity of 
\begin{equation}
   \mathcal{N}\left[\rho_N(t_X, t_B)\right] = \frac12
\end{equation}
for all times $t_X, t_B$. Hence, we see that maximal entanglement is preserved despite the spin flips.
 If we take into account that in realistic experiments we can only determine the time difference between the detection events with a resolution of $\Delta T$, we have to evaluate the negativity of the averaged density matrix
\begin{equation}
    \rho_{AV}(t,\Delta T) = \frac{1}{\Delta T}\int_t^{t+\Delta T} \text{d}t \rho_N(t, 0)
\end{equation}
where we set the time of detection of the biexciton photon to zero. The evaluation of the negativity yields
\begin{equation}
    \mathcal{N}\left[\rho_{AV}(t,\Delta T) \right] = \frac12 \left|\text{sinc}\left(\Omega \Delta T \right) \right|.
\end{equation}

\begin{figure}[htbp]
\centerline{\includegraphics[width=\linewidth]{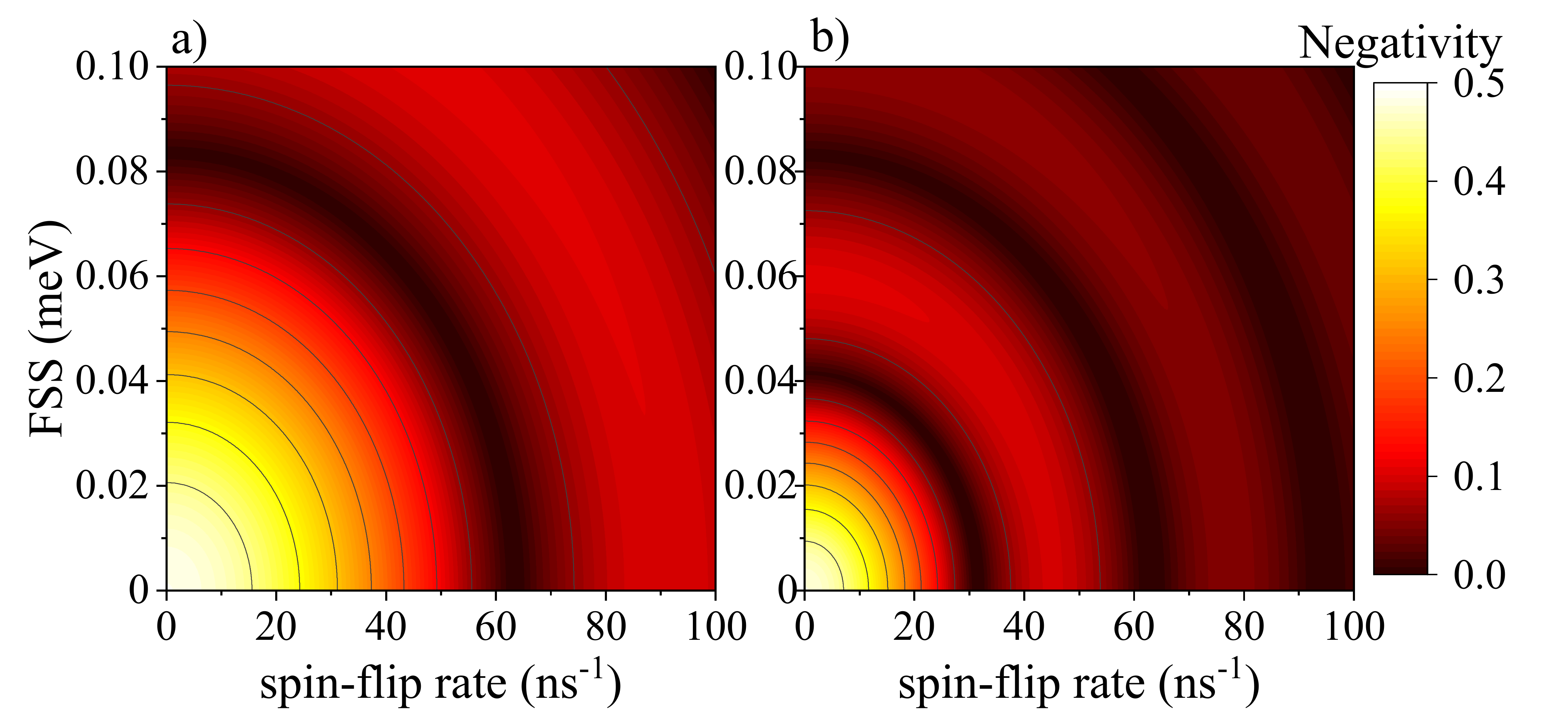}}
\caption{\footnotesize Negativity as a function of the spin-flip rate and the FSS for a resolution of (a) 50 ps and of (b) 100 ps. }
\label{fig:fig4}
\end{figure}

Fig.~\ref{fig:fig4} shows the negativity as a function of the spin-flip rate and the FSS for a given temporal resolution. This graph gives an overview of the negativity which can be expected depending on the FSS and spin-flip rates characteristic for QDs for two different resolutions (Fig.~\ref{fig:fig4} (a) for a 50 ps resolution and (b) for a 100 ps resolution). The overall entanglement quality is degraded as the FSS and the spin-flip rate are increased. This degradation is modulated by oscillations with respect to the FSS and the spin-flip rate. As can be deduced from the equation above, the temporal resolution of the detectors defines the frequency of these oscillations.

Perfect temporal resolution, that is, $\Delta T=0$, leads to a negativity being equal to 1/2, which means that perfect entanglement can be measured, provided that no additional dephasing processes need to be taken into account. 

The findings show that the interaction of the excitonic spins with the nuclear spins acts as a unitary transformation of the eigenstates and is in no way dissipative. Therefore, in the case of a perfect time post-selection, this mechanism is not detrimental for the generation of entanglement from QDs.

\section{Conclusion}
We investigated the entanglement in QDs showing spin-flip processes. The precession of the excitonic spins has been experimentally evidenced through the observation of anomalous oscillations in the rectilinear basis correlations. The associated quantum tomography measurement allows for studying the impact of spin-flip processes on the entanglement. 
We determined and studied the entanglement quality which can be ideally expected from a QD suffering from spin flips for a given temporal resolution. The theoretical model shows that the spin-flip rate acts the same way as a non-zero FSS and that a perfect temporal resolution would allow for ideal entanglement independent of the spin-flip rate. Temporal post-selection is therefore effective at providing perfect entanglement even in the presence of coherent processes modifying the eigenstates of the system since the latter are a coherent superposition of the excitonic states. Non-coherent processes affecting the excitonic and biexcitonic phases differently are the last remaining phenomena which cannot be effectively suppressed by post-selection. A coherent population of the biexciton and its coherent control through two-photon excitation or pulse-echo techniques~\cite{jayakumar} are in this case relevant to complement it.

The research leading to these results has received funding from the European Research Council (ERC) under the European Union’s Seventh Framework ERC Grant Agreement No. 615613 and from the German Research Foundation via project No. RE2974/12-1. We also acknowledge support from the Deutsche Forschungsgemeinschaft (DFG) through SFB
787. A. C. gratefully acknowledges support from the SFB 910: ``Control of self-organizing nonlinear systems".


\subsection{}
\subsubsection{}

\bibliographystyle{apsrev4-1}
\bibliography{qst_samir_neu}

\end{document}